\documentclass[twocolumn,twoside,slac_two]{revtex4}
\usepackage{graphicx}
\usepackage{fancyhdr}
\begin{document}

\title{The
Charm Quark Contribution to the Proton
Structure Function}

%
\author{M. Modarres}
\affiliation{Physics Department, Tehran University, 1439955961, Tehran, Iran}

\author{M. M. Yazdanpanah}
\affiliation{Physics Department, Shahid-Ba-Honar University, Kerman,
Iran  }

\begin{abstract}
The charm quark structure function $F^c_2$ and the longitudinal structure
function $F_l^p$ are directly sensitive to the gluon content of
proton and therefore are crucial in understanding of proton
structure function, in particular at low momentum transfer $Q^2$
and low Bjorken x. In the framework of perturbative QCD the charm
structure function is calculated in the leading order (LO) and
the proton structure function is investigated in the next leading
order (NLO) at small x region. The valence quark distribution is
obtained from the relativistic quark-exchange model. The
calculated $F_2^c(x,Q^2)$, $F_2^p(x,Q^2)$ and $F_L^p(x,Q^2)$, are
compared with the present available experimental data.
\end{abstract}

\maketitle

\thispagestyle{fancy}


\section{Valence Quark Distribution}
In the evaluation of valence quark
distribution, we assumed that the nucleon is composed of three
valence-quarks in the following way \cite{ref1}:
\begin{equation}
| {\alpha} {\rangle}={\cal N}^{{\alpha}^{\dagger}}| 0 {\rangle}
={1 \over \sqrt{3!}}{\cal N}^{\alpha}_{{\mu}_1 {\mu}_2 {\mu}_3}
q^{\dagger}_{{\mu}_1}q^{\dagger}_{{\mu}_2} q^{\dagger}_{{\mu}_3}
| 0 {\rangle},
\end{equation}
where $ \alpha$ designate the nucleon states ${\{\vec P}, M_S
,M_T \}$ and ${\mu}$ stand for the quark states ${\{\vec k} , m_s
, m_t , c\}$.  With the convention that there is a summation on
the repeated indices as well as integration over $\vec k$.
$q^{\dagger}$ $({\cal N}^{{\alpha}^{\dagger}})$ are the creation
operators for quarks (nucleons) and $ {\cal N}^{\alpha}_{{\mu}_1
{\mu}_2 {\mu}_3} $ are the totally antisymmetric nucleon wave
functions, i.e.
\begin{eqnarray}
{\cal N}^{\alpha}_{{\mu}_1 {\mu}_2 {\mu}_3}=
D(\mu_1,\mu_2,\mu_3;\alpha_i) \times
{ \delta({\vec k_1}+{\vec k_2}+{\vec k_3}-{\vec P}}) \nonumber\\
\times{\phi ({\vec k_1},{\vec k_2},{\vec k_3},{\vec P})}.
\end{eqnarray}
The $D(\mu_1,\mu_2,\mu_3;\alpha_i)$ depend on the Clebsch-Gordon
coefficients $C^{{j_1} {j_2} {j}}_{{m_1} {m_2} {m}}$ and the
color factor $\epsilon_{{c_1}{c_2}{c_3}}$,
\begin{widetext}
\begin{equation}
D(\mu_1,\mu_2,\mu_3;\alpha_i) = {1 \over \sqrt{3!}}
\epsilon_{c_1c_2c_3} {1 \over \sqrt 2}\sum_{s,t=0,1} C^{{1\over
2} s {1 \over 2}}_{m_{s_\sigma} m_s M_{S_{\alpha_i}}} 
C^{{1 \over 2} { 1 \over 2}  s}_{m_{s_\mu} m_{s_\nu} m_s} 
C^{{1 \over 2} t {1 \over 2}}_{m_{t_\sigma} m_t M_{T_{\alpha_i}}} C^{{1
\over 2}{ 1 \over 2} t}_{m_{t_\mu} m_{t_\nu} m_t}
\end{equation}
\end{widetext}
The ${\phi ({\vec k_1},{\vec k_2},{\vec k_3},{\vec P})}$ are the
nucleon wave functions in terms of quarks and we write it in a
Gaussian form (b $\simeq$ nucleon radius) :
\begin{eqnarray}
\phi ({\vec k_1},{\vec k_2},{\vec k_3},{\vec P})= {\left(\frac
{3b^4}{{\pi}^2}\right)^{3 \over 4}} \times~~~~~~~~~~~~~~~~~~~~~~~~\nonumber\\
\exp[-b^2(\frac {(k_1^2+k_2^2+k_3^2)}{2}+\frac {{b^2}{P^2}}{6}].
\end{eqnarray}

We can define the nucleus state based on nucleon creation
operators, i.e.
\begin{equation}
|{\cal A}_i=3 {\rangle}={(3!)^{-{1 \over 2}}}
{\chi}^{{\alpha}_1{\alpha}_2{\alpha}_3}{\cal
N}^{{{\alpha}_1}^{\dagger}} {\cal N}^{{{\alpha}_2}^{\dagger}}{\cal
N}^{{{\alpha}_3}^{\dagger}}|0{\rangle},
\end{equation}
where ${\chi}^{{\alpha}_1{\alpha}_2{\alpha}_3}$ are the complete
antisymmetric nuclear wave functions (they are taken from Faddeev
calculations with Reid soft core potential.

The quark momentum distributions with fixed flavour in a three
nucleon system are defined as,
\begin{equation}
{\rho}_{\bar {\mu}}(\vec k;{\cal A}_i)={{\langle} {\cal A}_i=3
|q^{\dagger}_{\bar {\mu}}q_{\bar {\mu}} |{\cal A}_i=3{\rangle}
\over {\langle}{\cal A}_i=3 |{\cal A}_i=3{\rangle}}~~~~~,
\end{equation}
where the sign bar means no summation on $ m_t$ and integration
over $\vec k$ in the $\mu$ indices. By using the above definition,
we can calculate the quark momentum distribution for each flavour.
In the above equation we use, $\chi(x,y,cos\theta)$,  the Fourier
transform of the nucleus wave function.

By considering the relativistic correction, the valence parton
distribution at each $Q^2$ can be related to momentum
distribution for each flavour according to the following equation,
\begin{eqnarray}
 q^v(x,Q^2;{\cal A}_i)={1 \over (1-x)^2}\int \rho_q({\vec
k};{\cal A}_i) \times ~~~~~~~~~~~~~~~ \nonumber\\
\delta({x \over (1-x)}-{k_+ \over M}) \ d {\vec k}.
\end{eqnarray}
Evaluating the angular integration, we find that,
\begin{equation}
q^v(x,Q^2;{\cal A}_i)={{2\pi M} \over {(1-x)^2}}
{\int_{k_{min}}^{\infty}} \rho_q({\vec k};{\cal A}_i)k \ dk
\end{equation}
with
\begin{equation}
k_{min}(x)={{({{xM} \over {1-x}}+\epsilon_0)^2-m^2} \over
{2({{xM} \over {1-x}}+\epsilon_0)}},
\end{equation}
where m (M) is the quark (nucleon) mass , $k_+$ is the light-cone
momentum of initial quark and $\epsilon_0$ is the quark binding
energy.  The valence quark distribution of a bound nucleon can be
derived from the free nucleon valence quark distribution function
by using the convolution approximation,
\begin{equation}
q^v(x,Q^2;{\cal A}_i)=\sum_N \int q^v ({x \over y_{{\cal
A}_i}},Q^2;N)f_{N/{\cal A}_i}(y_{{\cal A}_i}) \ d{y_{{\cal A}_i}},
\end{equation}
where $f_{N/{\cal A}_i}(y_{{\cal A}_i})$ are the nucleon momentum
distributions in the nucleus. By taking into account the fact that
$f_{N/{\cal A}_i}(y_{{\cal A}_i})$ are large only around $x \over
{\langle y_{{\cal A}_i} \rangle}$ we can write [23]
\begin{equation}
\Delta q^v({x \over {\langle y_{{\cal A}_i} \rangle}},Q^2;N)=
\Delta q^v(x,Q^2;{\cal A}_i)
\end{equation}
with ${\langle y_{{\cal A}_i}\rangle}=1+{{\bar \epsilon} \over M}$
and $\bar \epsilon$ being the average removal energy of the
nucleon. A typical ansatz for the parton distribution is usually
parameterized as,
\begin{equation}
x~ {\cal P}(x,Q^2)=A_{\cal P}{\eta_{\cal P}}x^{a_{\cal P}}(1-
x)^{b_{\cal P}} (1+\gamma_{\cal P} x+\varrho_{\cal P} x^{1 \over
2})
\end{equation}
where $A_{\cal P}$ is the normalization factor. 

\section{NLO Evolution Procedure}
It is appropriate to use the Mellin and inverse Mellin transformation to
calculate the NLO parton distributions in the $(x,Q^2)$-plane \cite{ref1}.

In the NLO, $F^p_2(x,Q^2)$ is related to the quark, antiquark and
gluon distributions, as follows:
\begin{widetext}
\begin{equation}
F^p_2(x;Q^2)= x \sum_{q=u,d,s} {e_q^2} \{q(x,Q^2)+ {\bar
q}(x,Q^2)\}
+{\alpha_s(Q^2) \over 2 \pi} [C_q \otimes ({ q(x,Q^2)+ {\bar
q}^N(x,Q^2}) +{2} C_g \otimes G(x,Q^2)]+ F_2^c(x,Q^2,m_c^2),
\end{equation}
\end{widetext}
where $\otimes$ means the convolution and it is defined as,
\begin{equation}
 C_{\cal P} \otimes {\cal P}=\int_x^1 {\ dy \over y} C_{\cal
P}({x \over y}) { \cal P}(y, Q^2).
\end{equation}

The charm quark contribution to the proton structure function
$F_2^c(x,Q^2,m_c^2) $ has the following form in the LO limit, if
${1\over x}\ge 1+({2m_c\over Q})^2$, (note that for small x, with
this condition  $Q^2$ can become less than  $m^2_c$, 
\begin{eqnarray}
F_2^c(x,Q^2,m_c^2)=2xe^2_c{\alpha_s({\mu^{\prime}}^2) \over 2
\pi}\int_{ax}^1{dy\over y}C^c_{g}({x\over y},({m_c\over
Q})^2) \nonumber\\
\times g(y,{\mu^{\prime}}^2)~~~~~~~
\end{eqnarray}
where $a=1+4{m_c^2\over Q^2}$ and ($n_f$ is the number of active
flavours)
\begin{equation}
\alpha_s(Q^2)= {4\pi\over
\beta_0\ln(Q^2/\Lambda^2)}-{4\pi\beta_1\over
\beta_0^3}{\ln\ln(Q^2/\Lambda^2)\over \ln(Q^2/\Lambda^2)}
\end{equation}
with ($\Lambda_{\overline{MS}}=200 MeV$)
\begin{equation}
\beta_0=11-{2\over 3}n_f, \ \ \ \beta_1=102-{38\over 3}n_f.
\end{equation}
The longitudinal SF,$F^p_L(x;Q^2)$ is obtained as:
\begin{widetext}
\begin{equation}
F^p_L(x;Q^2)= x{\alpha_s(Q^2) \over 2 \pi} \sum_{q=u,d,s}
[C_{q,L}\otimes ({ q(x,Q^2)+ {\bar q}^N(x,Q^2}) +{2} C_{g,L}
\otimes G(x,Q^2)]
+ F_L^c(x,Q^2,m_c^2)
\end{equation}
\end{widetext}
with
\begin{equation}
C_{q,L}={8\over 3}z, \ \  C_{g,L}=2z(z-1).
\end{equation}
We use the LO expression for $F_L^c(x,Q^2,m_c^2)$, which is
the same as equation (3) i.e.
\begin{eqnarray}
F_L^c(x,Q^2,m_c^2)=2xe^2_c{\alpha_s({\mu^{\prime}}^2) \over 2
\pi}\int_{ax}^1{dy\over y}C^c_{g,L}({x\over y},({m_c\over
Q})^2) \nonumber\\
\times g(y,{\mu^{\prime}}^2)~~~~~~~
\end{eqnarray}

Here we assume the SU(3) flavour-symmetric sea quark
distributions ${ \bar q}={ \bar u}={ \bar d}={ \bar s}={ s}$. In
addition we consider the sea quark and gluon contributions to
vanish in the static point $\mu_0^2\ll Q^2$, ( we set  $\mu_0^2 =
0.32 GeV^2$),i.e,
\begin{equation}
G(x,\mu_0^2)=0~~~~~~~~~~~~~~~\bar q(x,\mu_0^2)=0.
\end{equation}
Note that, this is the scale where the above initial condition
is approximately satisfied i.e. the $\mu_0^2$ scale is
determined from the intermediate $Q^2_0$ scale by evolving
downward the second moment of the valence quark distributions
such that the gluon and sea quark distributions are approximately
zero at the $\mu_0^2$ scale.
\begin{figure}
\includegraphics[width=65mm]{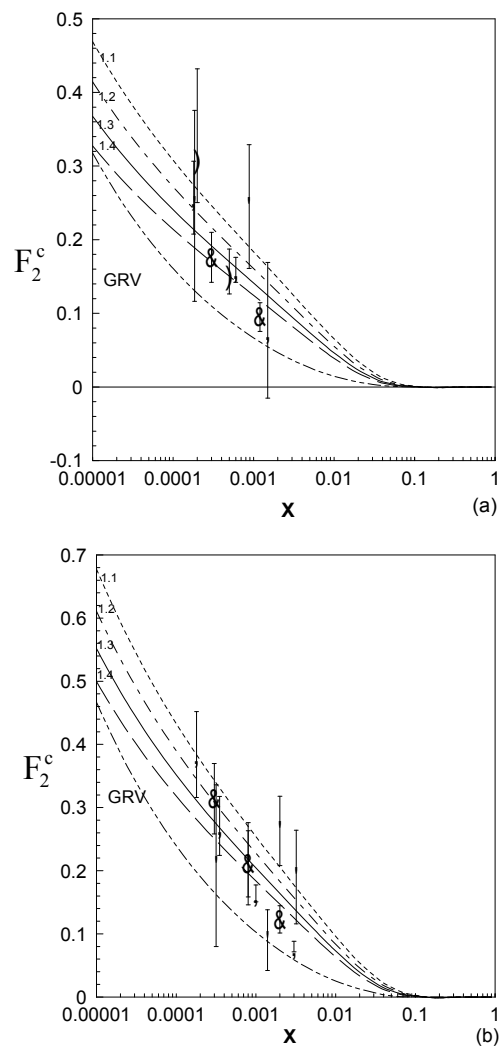}
\caption{The charm
contribution to the proton structure function at (a) $Q^2=6 GeV^2$
and (b) $Q^2=10 GeV^2$  for various charm mass. The data are from various
ZEUS  and H1 collaborations experiments.}
\end{figure}

\section{Results and discussions}

In figure 1, we present the charm quark contribution to the Structure function of
proton at ( $Q^2=6 GeV^2$)  with different
charm mass values i.e. $m_c=1.1 GeV$ (small-dash curve),
$m_c=1.2 GeV$ (dash-small-dash curve), $m_c=1.3 GeV$( full curve)
and $m_c=1.4 GeV$ (dash curve). The data are those of H1 \cite{ref2} and
ZEUS collaborations \cite{ref3}, i.e. the squares (ZEUS,1997), circles
(H1,2000), diamonds (ZEUS,2000) and triangles (ZEUS,2004). Our
results are in very good agreement with the present available
data. By reducing the charm mass, the charm structure function of
proton increases but it still passes though the  data. It also
becomes zero for $x>0.1$. Obviously the structure function
becomes larger as we increase $Q^2$. The calculated
$F_2^c(x,Q^2)$ by using the gluon distribution of GRV (heavy-full
curves) with $m_c=1.3$ shows less charm quark contribution to
the structure function of proton.

\begin{figure}
\includegraphics[width=65mm]{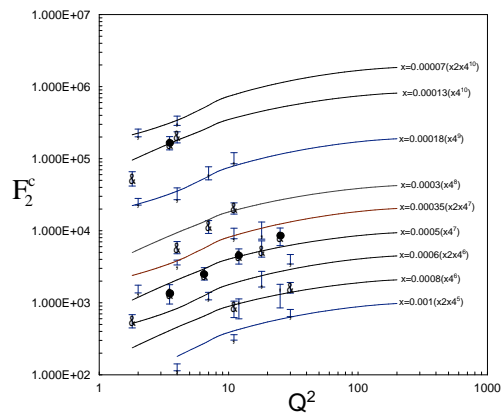}
\caption{The $Q^2(GeV^2)$ dependence of charm contribution to SF of
proton for the various x values. The data are from  ZEUS [2] and
H1 [3] collaboration experiments.}
\end{figure}
\begin{figure}
\includegraphics[width=70mm]{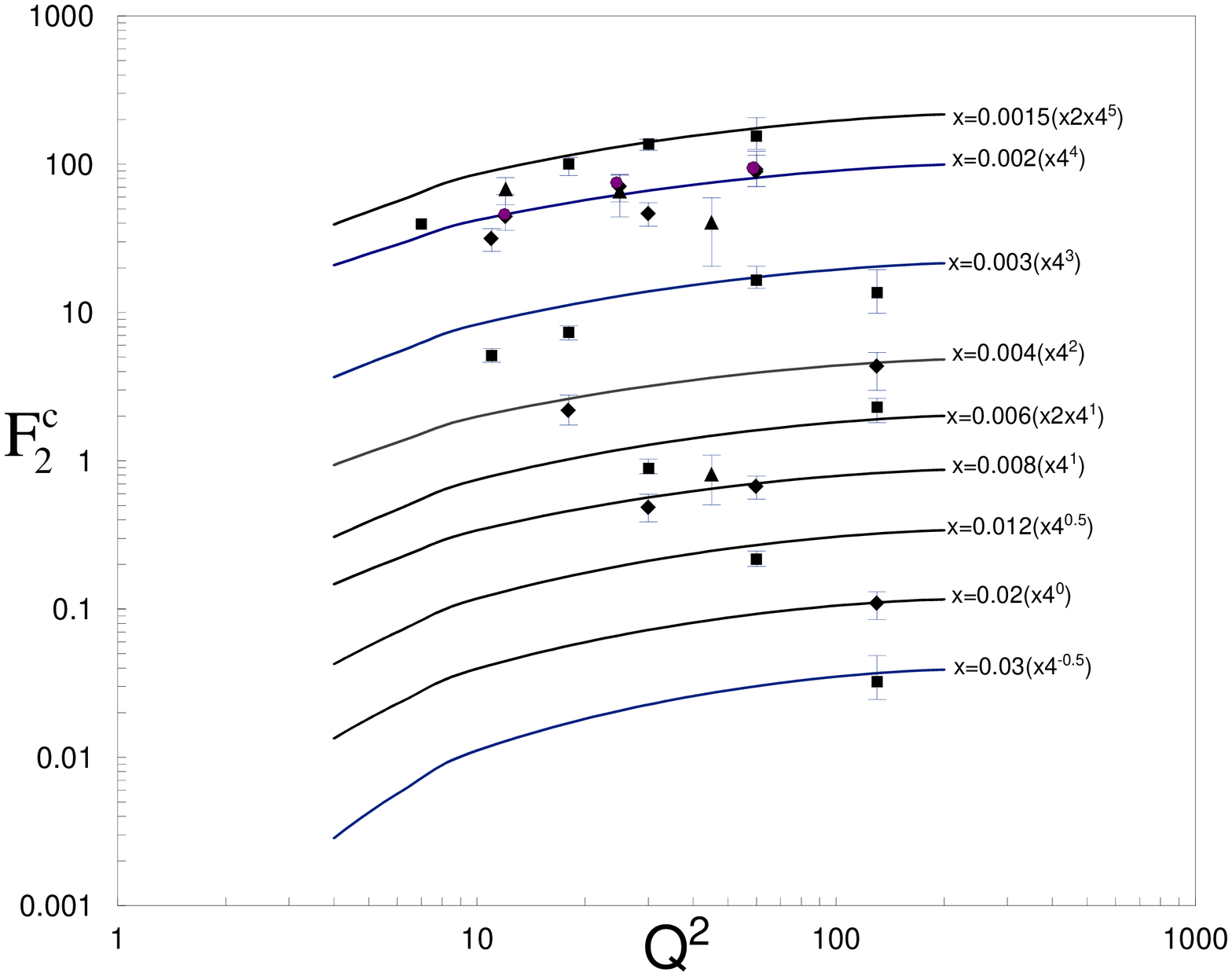}
\caption{As figure. 2. }
\end{figure}

Figures 2 and 3 show the comparison of  $Q^2$ dependence of
charm contribution to the structure function of proton for various x values with
the available  ZEUS data (the triangles (1997), circles (2000) and
squares (2004)). We get  a reasonable result with respect to the
data [2].

The $Q^2$  dependence of longitudinal structure function of proton are given in
figure 4. The data are from  H1 collaboration
experiments: the circles (H1,2001) and triangles (H1,1996).
Again there are a good agreement between our calculated results and
the experimental predictions data. These show that our gluon
distribution can reasonably well treat the PGF process.
\begin{figure}
\includegraphics[width=60mm]{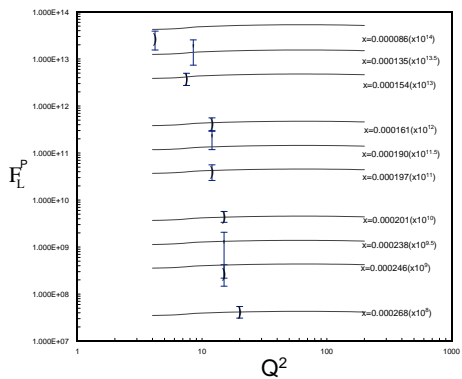}
\caption{The $Q^2(GeV^2)$ dependence of longitudinal SF of Proton. The data
are from  H1 collaboration experiments.}
\end{figure}
We have observed a linear $Q^2$-dependence in the unpolarized
structure function which is in a very good agreement with the
available data. A smooth behavior was found for the sea-quark and
gluon distribution functions. The agreement between our uniquely
gluon and sea quarks distributions and the available data are
encouraging. In particular we fully reproduced the gluon
distribution. We note that the above  assumption
where the gluon and the sea-quark distributions can be generated
entirely radiatively from valence quark may not be valid,
especially at small x.

In summary, by using our recent complete NLO calculation in the
conventional $\overline{MS}$  factorization scheme, we have
updated our previous NLO results by including the charm
contribution to the  SF of proton. We have found that  the proton
structure function has approximately the same $Q^2$-dependence as
the data and the whole results are consistent with the available
experiments. Our calculation shows a similar scaling violation as
the one observed in the experiment for the small x. The idea that
at low x the scaling violation of $F^N_2(x,Q^2)$ comes from the
gluon density alone and does not depend on the quark density
  which was tested in our previous work, is
still valid with a good accuracy. So, as before, we may conclude
that the gluon is  the dominant source of the parton in the small
x region. However, it is well known that the theoretical
interpretation of SF is complicated at low $Q^2$. Since, in this
region the higher twist contributions which are proportional to
$Q^{-2}$ and $Q^{-4}$ are not included in  the DGLAP equations.


\begin{thebibliography}{9}   
\bibitem{ref1}M. M. Yazdanpanah and M. Modarres, Eur.Phys.J.A, 7
(2000) 573.\\
M. M. Yazdanpanah and. M. Modarres, Eur.Phys.J.A, 6 (1999) 91.
\bibitem{ref2}H1 collaboration, T. Ahmed et al, Nucl.Phys.B, 439 (1995)
471; S. Aid et al, Nucl.Phys.B, 445 (1996) 3; I. Abt et al.,
Nucl.Phys., B407 (1993) 515; C. Adloff et al , Eur.Phys.J.C, 19
(2001) 269; Eur. Phys. J. C21 (2001) 33; Eur. Phys.J.C, 13 (2000)
609;  A. Aktas et al., Phys.Lett.B, 598 (2004) 159 and Phys.Lett.
B407 (2002) 402.
\bibitem{ref3}ZEUS collaboration, M. Derrick et al., Z.Phys.C, 72 (1996) 399;
M. Derrick et al., Phys.Lett.B, 316 (1993) 412 and Phys.Lett.B,
407 (1997) 402; J. Breitwag et al., Euro.Phys.J.C 12 (2000) 35 and
Eur.Phys.J.C, 7 (1999) 609; S. Chekanov et al., Phys.Rev.D, 69
(2004) 012004; S.Chekanov et al, Eur.Phys.J.C, 21 (2001) 443.

\end{thebibliography}
\end{document}